\documentclass[twocolumn]{aastex631}

\submitjournal{ApJ}
\shorttitle{X-ray Radiative Transfer Modeling of CAL87}
\shortauthors{Tsujimoto et al.}
\graphicspath{{./}}

\begin{document}

\title{Spectral Modeling of the Supersoft X-ray Source CAL87 based on Radiative Transfer Codes}
\correspondingauthor{Masahiro Tsujimoto}
\email{tsujimot@astro.isas.jaxa.jp}

\author[0000-0002-9184-5556]{Masahiro Tsujimoto}
\affil{Institute of Space and Astronautical Science (ISAS), Japan Aerospace Exploration Agency (JAXA),\\3-1-1 Yoshinodai, Chuo-ku, Sagamihara, Kanagawa 252-5210, Japan}

\author[0000-0003-2161-0361]{Misaki Mizumoto}
\affil{Faculty of Education, University of Teacher Education Fukuoka, 1-1 Akama-bunkyo-machi, Munakata, Fukuoka 811-4192, Japan}

\author[0000-0002-5352-7178]{Ken Ebisawa}
\affil{Institute of Space and Astronautical Science (ISAS), Japan Aerospace Exploration Agency (JAXA),\\3-1-1 Yoshinodai, Chuo-ku, Sagamihara, Kanagawa 252-5210, Japan}
\affil{Department of Astronomy, Graduate School of Science, The University of Tokyo, 7-3-1 Hongo, Bunkyo-ku, Tokyo 113-0033, Japan}

\author[0000-0003-2670-6936]{Hirokazu Odaka}
\affil{Department of Earth and Space Science, Graduate School of Science, Osaka University,\\1-1 Machikaneyama, Toyonaka, Osaka 560-0043, Japan}
\affil{Kavil IPMU (WPI), The University of Tokyo, 5-1-5 Kashiwanoha, Kashiwa, Chiba 277-8583, Japan}

\author{Qazuya Wada}
\altaffiliation{Regrit Partners, Inc., 3-2-9 K\={o}ji-machi, Chiyoda-ku, Tokyo, 102-0083, Japan}
\affil{Institute of Space and Astronautical Science (ISAS), Japan Aerospace Exploration Agency (JAXA),\\3-1-1 Yoshinodai, Chuo-ku, Sagamihara, Kanagawa 252-5210, Japan}
\affil{Department of Astronomy, Graduate School of Science, The University of Tokyo, 7-3-1 Hongo, Bunkyo-ku, Tokyo 113-0033, Japan}

\begin{abstract} 
 Super Soft X-ray Sources (SSS) are white dwarf (WD) binaries that radiate almost
 entirely below $\sim$1~keV. Their X-ray spectra are often complex when viewed with the
 X-ray grating spectrometers, where numerous emission and absorption features are
 intermingled and hard to separate. The absorption features are mostly from the WD
 atmosphere, for which radiative transfer models have been constructed. The emission
 features are from the corona surrounding the WD atmosphere, in which incident emission
 from the WD surface is reprocessed. Modeling the corona requires different solvers and
 assumptions for the radiative transfer, which is yet to be achieved. We chose CAL87, a
 SSS in the Large Magellanic Cloud, which exhibits emission-dominated spectra from the
 corona as the WD atmosphere emission is assumed to be completely blocked by the
 accretion disk. We constructed a radiative transfer model for the corona using the two
 radiative transfer codes; \texttt{xstar} for a one-dimensional two-stream solver and
 \texttt{MONACO} for a three-dimensional Monte Carlo solver. We identified their
 differences and limitations in comparison to the spectra taken with the Reflection
 Grating Spectrometer onboard the \textit{XMM-Newton} satellite. We finally obtained a
 sufficiently good spectral model of CAL87 based on the radiative transfer of the corona
 plus an additional collisionally ionized plasma. In the coming X-ray
 microcalorimeter era, it will be required to interpret spectra based on radiative
 transfer in a wider range of sources than what is presented here.
\end{abstract}


\section{Introduction}\label{s1}
Super Soft X-ray Sources (SSS) represent a particular class of the white dwarf (WD)
binary systems characterized by their extremely soft X-ray emission below
$\sim$1~keV. The X-ray luminosity reaches up to $\sim$$10^{38}$~erg~s$^{-1}$
\citep{greiner1996} close to the Eddington limit for a solar-mass WD. They are observed
either as a persistent SSS with steady nuclear burning on the WD surface or a transient
source observed in the SSS phase of classical novae. SSS are important for our
understanding of binary evolution and supernova origins.

Early low-resolution X-ray spectra of SSS roughly resemble the blackbody radiation with
a temperature of 20--100~eV. However, when observed with high-resolution X-ray grating
spectrometers, SSS exhibit numerous emission and absorption features over continuum
emission (e.g., \citealt{paerels01}). It is considered that the continuum emission with
absorption features are emitted from the WD atmosphere, while the emission lines are
produced in a hot tenuous corona surrounding the WD (e.g., \citealt{ebisawa2010,
ness2013}). In the corona, incident photons from the WD atmosphere are reprocessed.

\citet{ness2013} conducted a systematic study of high-resolution X-ray spectra of SSS
and found that they are categorized into two types; those mainly dominated by absorption
features (SSa) and others by emission features (SSe). \citet{ness2013} proposed an
interpretation that the X-ray spectra of SSa and SSe respectively have a lower and
higher fraction of the corona emission with respect to the WD surface emission. The
difference is caused by different degrees of the obscuration of the WD surface emission
by the companion star or the accretion disk. \citet{ness2013} further pointed out that
SSS with high inclination angles ($\gtrsim70^\circ$) tends to have the SSe spectra. This
interpretation is reinforced in the partially eclipsing system V5116 Sgr, in which the
SSa spectrum is seen out of the eclipse and the SSe spectrum is seen during the eclipse
\citep{sala2008,sala2010}.

\medskip

The next obvious step is to interpret the observed spectra with physical models and
constrain their physical quantities. This has been explored intensively for SSa
and was applied to explain some aspects of the observed data. \citet{lanz05} extended
an industry-standard radiative transfer code for the stellar atmosphere \texttt{TLUSTY}
into the WD regime and explained the CAL 83 spectra. \citet{rauch2010} developed another
model called \texttt{TMAP} (T\"{u}bingen Model Atmosphere Package) and fitted the SSS
phase spectra of the nova V4743 Sgr. Both models solve the radiative transfer for the
non-local thermal equilibrium (NLTE) condition under the plane-parallel geometry and
hydrostatic density distribution using the accelerated lambda iteration
solver. \cite{vanrossum2010} further explained the V2491 Cyg spectra using the
\texttt{PHOENIX} code calculating the WD atmosphere with an expanding envelope.

The SSe spectra, on the contrary, are less explored because of the increased
complexity. In addition to the WD atmosphere model, we need to construct the corona
model. The corona is photo-ionized by the WD atmosphere emission (hereafter called
``incident emission''). Reprocessed photons are emitted through radiative de-excitation
and recombination (``diffuse emission'') and electron scattering of the incident photons
(``scattered emission''). NLTE radiative transfer calculation is undoubtedly needed, but
different solvers and assumptions may be required from the ones used for the WD
atmosphere models.

\medskip

The purpose of this paper is to construct a physical model for SSe. We choose CAL87, a
representative SSe source, which is simple enough so that we can constrain the system
parameters based only on analytical calculations. We construct corona models using two
different codes using different solvers and compare their results with observations and
discuss their differences and limitations.

The structure of this paper is as follows. We describe the data in \S~\ref{s2} and the
model in \S~\ref{s3}. In the data section (\S~\ref{s2}), we give properties of the
target (\S~\ref{s2-1}), the observation and data reduction (\S~\ref{s2-2}), and the
characterization of the spectra using phenomenological models (\S~\ref{s2-3}). In the
model section (\S~\ref{s3}), we set up the assumption for the geometry and the
parameterization (\S~\ref{s3-1}), based on which we give an analytical estimation of the
parameter ranges (\S~\ref{s3-2}) and numerical verification of the parameter values by
comparing the data and the synthesized spectra (\S~\ref{s3-3}). In \S~\ref{s4}, we
construct physically motivated models using two different codes: one is \texttt{xstar}
using a one-dimensional (1D) two-steam solver (\S~\ref{s4-1}) and the other is
\texttt{MONACO} using a three-dimensional (3D) Monte Carlo solver (\S~\ref{s4-2}) of the
radiative transfer.  We identify discrepancies between the two models against the
observation and discuss their possible origins in \S~\ref{s5}. The quoted errors are
1$\sigma$ statistical uncertainty throughout the paper.

\section{Data}\label{s2}
\subsection{Target}\label{s2-1}
CAL87 \citep{long81} is a reasonably bright source located in the Large Magellanic Cloud
(LMC) at a distance of 48.1 $\pm$ 2.9~kpc \citep{macri06}. It is an eclipsing binary
with an orbital period of 10.6~hr \citep{pakull1988,callanan89,cowley90,schmidtke93},
suggesting a high inclination angle \citep{schandl1997}. The X-ray emission is
considered to come from the extended accretion disk corona to account for the shallow
and long X-ray eclipse due to the companion star \citep{schmidtke93,asai1998,ebisawa01}
and for the lack of changes in the X-ray emission line intensity in and out of the
eclipse \citep{ribeiro2014}.

\citet{ebisawa01} proposed a picture, in which the X-ray emission from the WD atmosphere
is permanently blocked, but its electron-scattered emission in the corona is
observed. The observed \ion{O}{7} and \ion{O}{8} edges, recognized with the CCD spectral
resolution, are footprints of the incident WD atmosphere emission. The corona radius is
estimated as $\sim$5 $\times 10^{10}$~cm, which is $\sim$0.4 times the orbital
separation, and the inclination angle is $\sim$73~degree to account for the observed
X-ray light curve. Emission lines should be observed from the corona, which were not
recognized in the CCD spectra, but indeed found later with the improved energy
resolution of X-ray grating spectrometers
\citep{orio2004,greiner2004,ebisawa2010,ribeiro2014}. We follow the picture thus
established.

\subsection{Observations and Data reduction}\label{s2-2}
CAL87 was observed with the \textit{XMM-Newton} observatory \citep{jansen2001} on
2003-04-18 for 21.8 hours covering two orbital cycles (the observation number 0153250101).
We used the Reflection Grating Spectrometer (RGS; \citealt{denherder2001}) to exploit
its high-resolution X-ray spectra as well as the Metal Oxide Semiconductor (MOS;
\citealt{Turner2001}) type detector of the European Photon Imaging Camera (EPIC) to
constrain the continuum level.

The data were retrieved from the archive and reduced using the Science Analysis System
(SAS) software version 13.5.0 with the calibration files available as of December 5,
2014. For the spectral fitting, we used \texttt{Xspec} version 12.12.1. We used the
standard processing using the \texttt{rgsproc} task to extract the dispersion of the
first order in the RGS data both for the source and background spectra. We combined the
RGS1 and RGS2 spectra using \texttt{rgscombine} task to improve the statistics for all
the data. The MOS data were processed with the standard processing using the
\texttt{emproc} task. We selected the fiducial events with a PATTERN of 0--12 and
removed those affected by the background flares. Then, we extracted the source and
background events respectively from a circle centered at the source with a radius
$40^{\prime\prime}$ and an annulus of the inner and outer radius of $75^{\prime\prime}$
and $100^{\prime\prime}$. Relative normalization between MOS1 and MOS2 was assumed
identical, while that between MOS and RGS was determined in the spectral fitting. The
spectra are averaged over the exposure time because the spectral shape does not change
in and out of the eclipse \citep{ebisawa2010}.

\subsection{Spectral characterization}\label{s2-3}
\begin{figure}
 \begin{center}
  \includegraphics[width=\columnwidth]{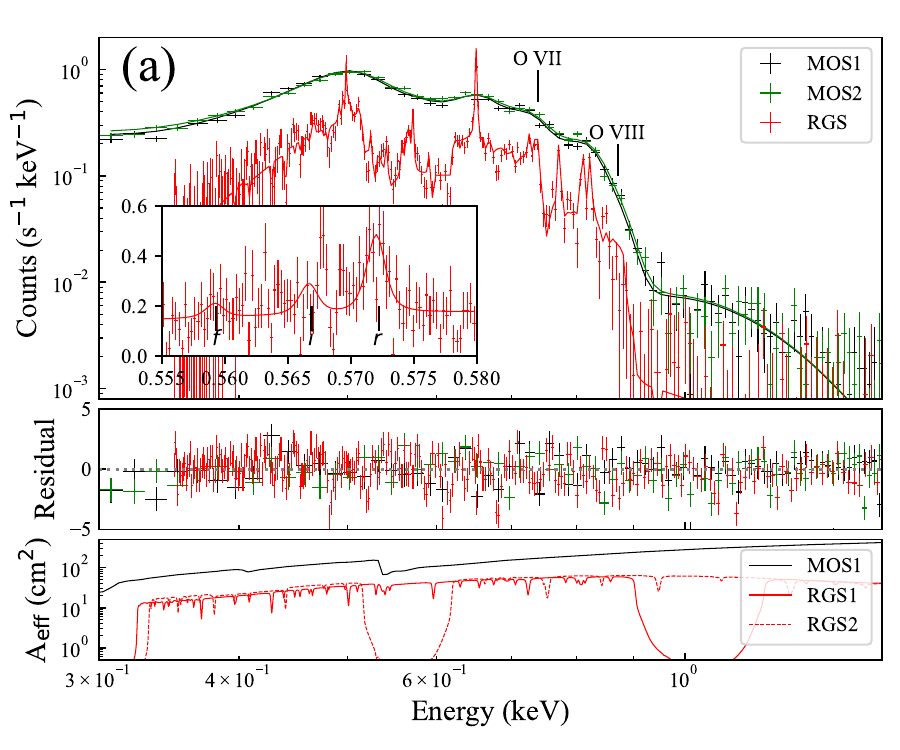}
  \includegraphics[width=\columnwidth]{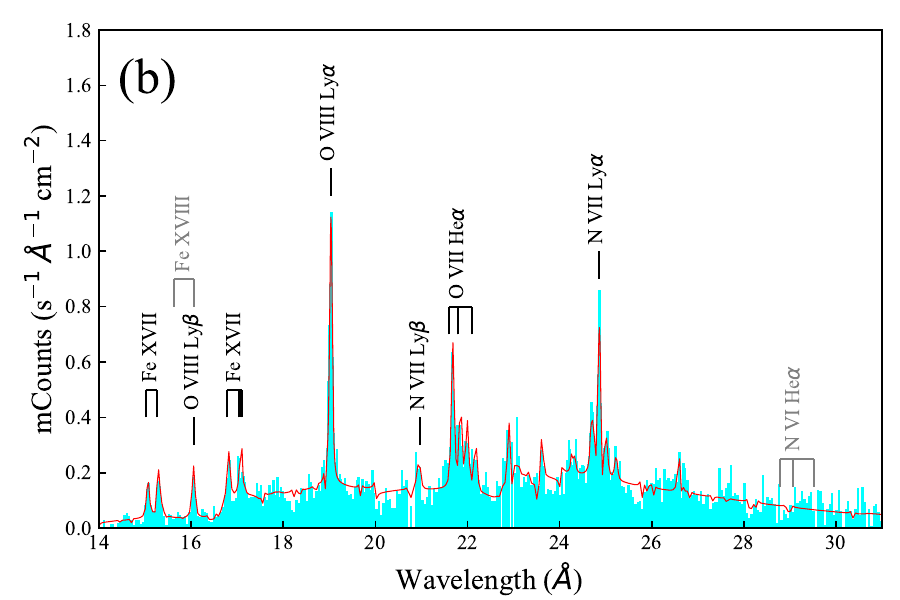} 
  \caption{Fitting results with a phenomenological model. (a) MOS1 (black), MOS2
  (green), and combined RGS (red) spectra in the 0.3--2.0~keV band. Two lines indicate
  the photoelectric absorption K edges by \ion{O}{7} and \ion{O}{8}. The upper panel
  shows the binned data with crosses and the best-fit models with solid lines. The inset
  shows a close-up view of the local fitting for \ion{O}{7} He$\alpha$ lines. The middle
  panel shows the residuals to the fit, and the bottom panel shows the effective areas of
  the individual detectors. One of the two RGS units (RGS2) does not cover the
  20.0--24.1~\AA\ range, where \ion{N}{6} Ly$\beta$ and \ion{O}{7} He$\alpha$ lines are
  located. 
  (b) The RGS spectrum in the 14--31~\AA~band (corresponding to $\sim$0.4--0.9~keV),
  where the vertical axis is in the linear scale. The identified emission lines are
  marked with vertical lines and labeled in black at their rest wavelengths.
  Those absent lines which would have been expected at slightly different
  ionization states (see Fig.~\ref{f03}) are also indicated in grey.  The cyan-shaded
  region and the red solid line indicate the data and the best-fit model, respectively. 
  }
 \label{f05}
 \end{center}
\end{figure}

The X-ray spectra of the two MOS and the combined RGS data are shown in Figure
\ref{f05}. We characterize them by constructing a phenomenological spectral model for
later use in physically-motivated models. For all the models in this work, we included the
attenuation by the Galactic interstellar medium (ISM) in the direction of CAL87 and the
ISM in the LMC using two photoelectric absorption models \citep[\texttt{tbabs} and
\texttt{tbvarabs}:][]{wilms00}. The column density of the former
($N_\mathrm{H}^\mathrm{Gal}$) was fixed at 7.58$\times$10$^{20}$~cm$^{-2}$
\citep{dickey1990} and that of the latter ($N_\mathrm{H}^\mathrm{LMC}$) was derived from
the fitting. The metal abundance for the LMC absorption was fixed to be half of the
Galactic ISM value \citep{Russell1992,Welty1999}.

The continuum model was constrained using the MOS spectra with the richer
statistics (Fig.~\ref{f05} a). The spectra are characterized by a sharp cut-off at
0.7--0.9~keV despite the monotonically increasing effective area of the telescope. We
thus used a single blackbody model with a photoelectric absorption by the \ion{O}{7} and
\ion{O}{8} K shell edges at 0.74 and 0.87~keV, respectively. The same model was used in
the previous work using X-ray CCD spectra \citep{asai1998,ebisawa01}. The bolometric luminosity
corrected for the interstellar absorption ($L_{\rm obs}$) is $1.4 \times
10^{37}$~erg~s$^{-1}$ assuming the spherical emission.

\begin{table}
 \begin{center}
  \caption{Best-fit parameters of the phenomenological model (continuum).}
  \label{t01}
  \begin{tabular}{llc}
   \hline\hline
   \multicolumn{2}{l}{Parameter} & Value \\
   \hline
   Galactic absorption & $N_\mathrm{H}^\mathrm{Gal.}$ ($10^{21}$~cm$^{-2}$) & 0.758 (fix) \\
   LMC absorption & $N_\mathrm{H}^\mathrm{LMC}$ ($10^{21}$~cm$^{-2}$) & 3.61$\pm0.02$ \\
   Blackbody & $kT_\mathrm{BB}$ (eV) & 80.1$^{+0.3}_{-0.2}$ \\
   & Norm. ($\times10^{-4}$) & $5.67\pm0.06$ \\
   Edge energy & $E_\mathrm{O\,VIII}$ (keV) & 0.885$\pm0.004$ \\
   & $E_\mathrm{O\,VII}$ (keV)  & 0.743$\pm0.002$ \\
   Absorption edge depth & $\tau_\mathrm{O\,VIII}$ & 2.59$_{-0.13}^{+0.27}$ \\
   & $\tau_\mathrm{O\,VII}$ & 1.22$_{-0.11}^{+0.05}$ \\
   Luminosity  &$L_\mathrm{bol}$ ($10^{37}$~erg~s$^{-1}$) & $1.39\pm0.17$ \\
   \hline
   \multicolumn{2}{l}{$\chi_\mathrm{red}^{2}$ (dof)} & 1.20 (1744) \\
   \hline
  \end{tabular}
 \end{center}
\end{table}

Upon the best-fit continuum model, we added line components by using both the MOS
and RGS (Fig.~\ref{f05} b). A handful of emission lines were identified in the
RGS spectrum. We used Gaussian models for each line. From the best-fit energies, we
identified emission lines of \ion{O}{8} Ly$\alpha$ and Ly$\beta$, \ion{O}{7} He$\alpha$,
\ion{N}{7} Ly$\alpha$ and Ly$\beta$, and four \ion{Fe}{17} lines for the transition to
the ground state (2s)$^{2}$(2p)$^{6}$ from the excited states shown in
Table~\ref{t03}. All of them are electric dipole transitions of a large oscillator
strength among the \ion{Fe}{17} lines \citep{chen2003}. These identifications are
reported in \citet{ebisawa2010}.

\begin{deluxetable}{clccc}
\tablecaption{\ion{Fe}{17} lines}
\label{t03}
\tablehead{
  \colhead{Label} &
  \colhead{Upper level\tablenotemark{a}} &
  \colhead{Energy} &
  \colhead{$gf$\tablenotemark{b}} &
  \colhead{$\epsilon$\tablenotemark{c}}\\
  \colhead{} &
  \colhead{} &
  \colhead{(keV)} &
  \colhead{} &
  \colhead{(cm$^{3}$~s$^{-1}$)}
}
\startdata
3C & (2s)$^{2}$(2p)$^{5}$(3d)$^{1}$ $^{1}P_{1}$ & 0.826 & 2.49 & 1.8$\times$10$^{-15}$ \\
3D & (2s)$^{2}$(2p)$^{5}$(3d)$^{1}$ $^{3}D_{1}$ & 0.812 & 0.64 & 0.6$\times$10$^{-15}$ \\
3F & (2s)$^{2}$(2p)$^{5}$(3s)$^{1}$ $^{3}P_{1}$ & 0.738 & 0.10 & 1.1$\times$10$^{-15}$ \\
3G & (2s)$^{2}$(2p)$^{5}$(3s)$^{1}$ $^{1}P_{1}$ & 0.726 & 0.13 & 1.5$\times$10$^{-15}$ \\
\enddata
\tablenotemark{a}{Lower level is the ground state (2s)$^{2}$(2p)$^{6}$ $^{1}S_{0}$ for all.}
\tablenotetext{b}{Weighted oscillator strength.}
\tablenotetext{c}{Emissivity for the collisionally ionized plasma at a temperature of
 0.6~keV \citep{smith01}.}
\end{deluxetable}

Based on the line identifications, we fitted individual line complexes in a narrow
(20--30~eV) energy range. For each of the Ly$\alpha$ and Ly$\beta$ line
complexes, we included two lines ($^{2}P_{1/2}$ and $^{2}P_{3/2}$). The relative energy
and intensity ratio between the two lines were fixed. The line intensity, energy shift,
and broadening were fitted collectively. For the He$\alpha$ line complex, we
modeled with three lines of resonance ($r$), inter-combination ($i$), and forbidden
($f$). For the \ion{Fe}{17} line complex at 0.73--0.74~keV, we modeled with two
lines (3F and 3G; Table~\ref{t03}). For the \ion{Fe}{17} line complex at
0.81--0.83~keV, we modeled with two lines (3C and 3D; Table~\ref{t03}). The relative
energy of each line was fixed. The line energy shift and broadening were fitted
collectively, while the line intensity was fitted individually.

\begin{figure}
 \begin{center}
  \includegraphics[width=\columnwidth]{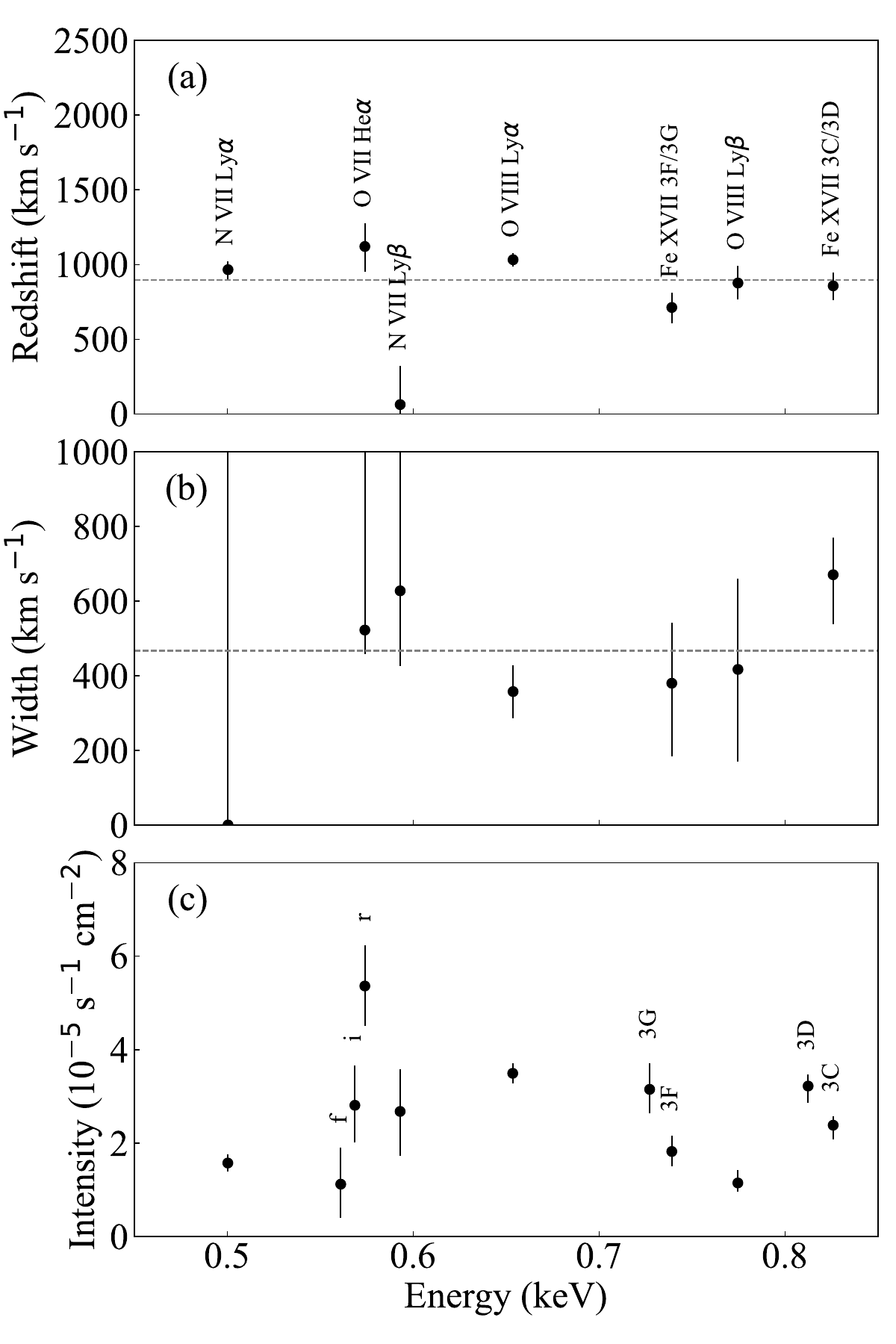}
  \caption{Results of the fitting of individual line complex for (a) energy shift, (b)
  line broadening (Gaussian standard deviation), and (c) intensity. The weighted mean
  values are shown with dashed lines in (a) and (b). The 1 $\sigma$ statistical error
  bar is given.  }
 \label{f10}
 \end{center}
\end{figure}

The result of the line complex fitting is shown in Figure~\ref{f10}. For the line shift
and broadening, most complexes exhibit a redward shift with a mean of
$\sim$900~km~s$^{-1}$ and broadening of $\sim$400~km~s$^{-1}$, which have been reported
in \citet{orio2004} and \citet{ribeiro2014}. The line shift of \ion{N}{7}
Ly$\beta$ deviates from the trend, which may be biased by the RGS chip gap
eliminating a large fraction of the line profile.

\section{Model}\label{s3}
\subsection{Assumptions}\label{s3-1}
\begin{figure}[htbp]
 \begin{center}
  \includegraphics[width=0.8\columnwidth]{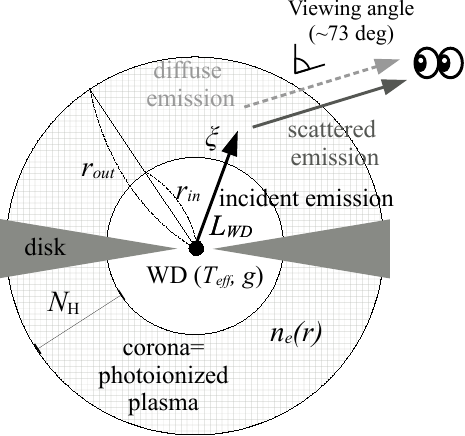}
 \end{center} 
 \caption{Setup of the geometry. The notations are given in the text.}
 \label{f01}
\end{figure}

We assume a static and spherically-symmetric geometry (Fig.~\ref{f01}), in which a
point-like source (WD) at the center is surrounded by the photoionized plasma
($=$corona).  The WD has an effective temperature of $T_{\mathrm{eff}}$, a surface
gravity of $g$, and a luminosity of $L_{\mathrm{WD}}$. It emits incident photons that
photo-ionize the corona. It is assumed that the observer does not directly
observe the incident emission from the WD being fully blocked by the disk but does
observe the diffuse emission from the corona and the scattered emission originating in
the WD. The inner and outer radii of the corona are $r_{\mathrm{in}}$ and
$r_{\mathrm{out}}$, respectively. The electron density profile is assumed to be
$n_{\mathrm{e}}(r) = n_{\mathrm{e,in}} (r/r_{\mathrm{in}})^{-2}$. The metal abundance in
the corona was fixed to 0.5 times the solar value \citep{wilms00}. The abundance would
be determined from the data after a successful spectral modeling but is left out of
scope for this study.

\subsection{Analytical estimation}\label{s3-2}
We first estimate the values of the model parameters analytically. The ionization parameter
of the corona at $r_{\mathrm{in}}$ is given by
\begin{eqnarray}
\label{e1}
 \xi = L_{\mathrm{WD}} / n_{\mathrm{e,in}} r_{\mathrm{in}}^2
\end{eqnarray}
in the unit of erg~s$^{-1}$~cm.  This is almost constant throughout the corona due to
the assumed radial dependence of $n_{e}(r)$. The electron column density
$N_{\mathrm{e}}$ in the line of sight is
\begin{eqnarray}
\label{e2}
 N_{\mathrm{e}} = \int_{r_{\mathrm{in}}}^{r_{\mathrm{out}}} n_{\mathrm{e}}(r) dr =
 n_{\mathrm{e,in}}r_{\mathrm{in}}^{2}\left(\frac{1}{r_{\mathrm{in}}} -
					   \frac{1}{r_{\mathrm{out}}}\right).
\end{eqnarray}
The electron scattering optical depth is given by 
\begin{eqnarray}
 \label{e3}
 \tau_{\mathrm{es}} = \sigma_{\mathrm{T}}N_{\mathrm{e}}, 
\end{eqnarray}
in which $\sigma_{\mathrm{T}}=6.65\times10^{-25}$~cm$^{2}$ is the Thomson
cross-section. We make the hydrogen and electron column densities equal as
$N_{\mathrm{H}}=N_{\mathrm{e}}$. Finally, the observed luminosity $L_{\mathrm{obs}}$ is
approximated as
\begin{eqnarray}
 \label{e4}
 L_{\mathrm{obs}} = (1-e^{-\tau_{\mathrm{es}}}) L_{\mathrm{WD}} 
\end{eqnarray}

\begin{figure}[htbp]
 \begin{center}
  \includegraphics[width=1.0\columnwidth, clip]{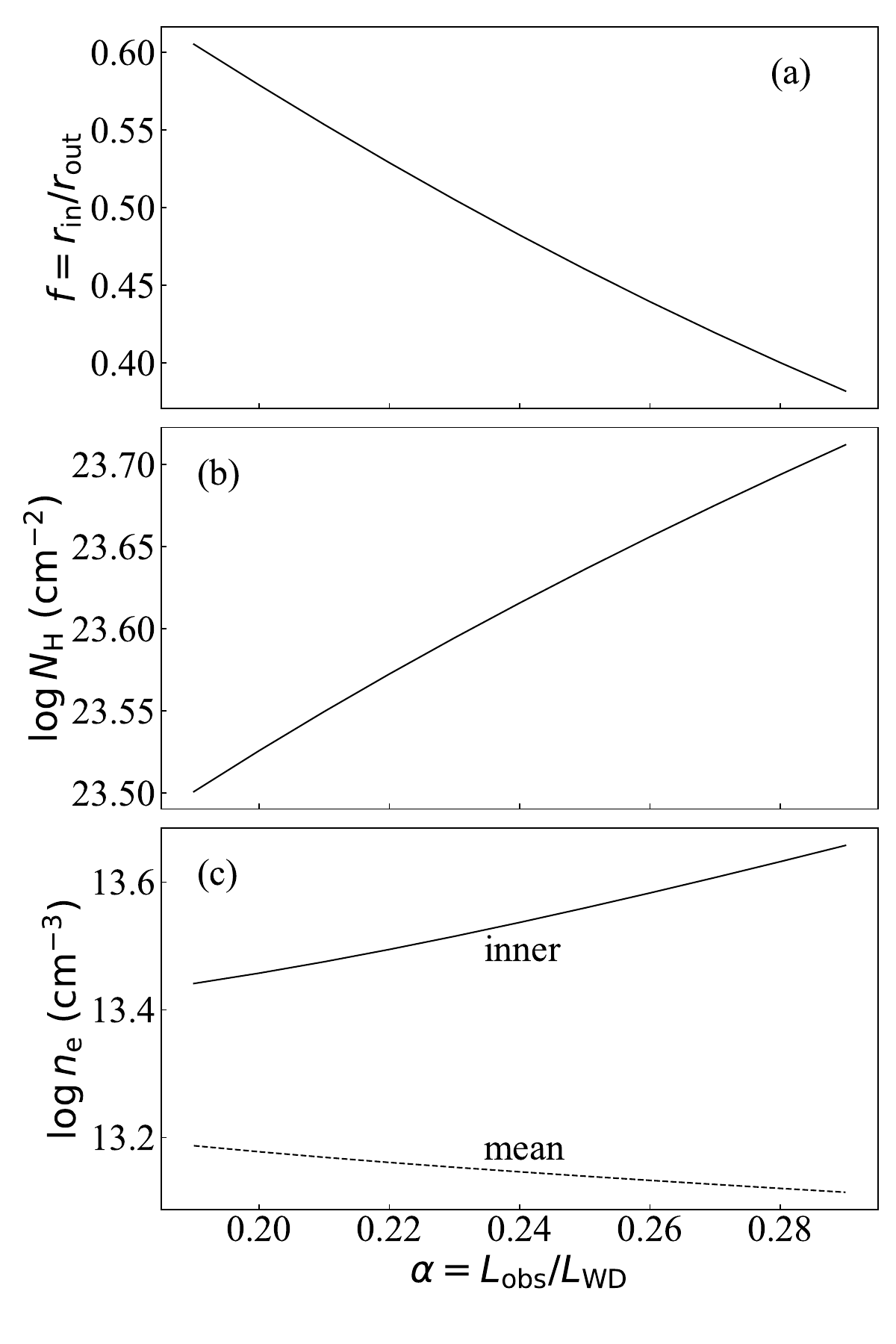}
 \end{center} 
 \caption{Analytical solution of (a) $f=r_{\mathrm{in}}/r_{\mathrm{out}}$, (b) column
 density through the corona $N_{\mathrm{H}}$, and (c) electron density at $r_{in}$
 (solid) and volume mean (dashed) as a function of $\alpha=L_{\mathrm{obs}}/L_{\mathrm{WD}}$.}
 \label{f02}
\end{figure}

We now substitute $L_{\mathrm{obs}}=1.4 \times 10^{37}$~erg~s$^{-1}$ (Table~\ref{t01}),
$r_{\mathrm{out}}= 4.8 \times 10^{10}$~cm from the partial eclipse in the X-ray light
curve \citep{ebisawa01}, and $\xi = 10^{2.5}$~erg~s$^{-1}$~cm from the observed lines,
which we will derive later using a numerical calculation in \S~\ref{s3-3-1}. The
unknown variable $L_\mathrm{WD}$ is parameterized by
$\alpha=L_\mathrm{obs}/L_\mathrm{WD}$, in which $ 0 \le \alpha \le 1$. As a function of
$\alpha$, we solve the other unknown variables $n_{\mathrm{e,in}}$, $N_{\mathrm{H}}$,
and $r_{\mathrm{in}}$ normalized by $r_{\mathrm{out}}$ as
$f=r_{\mathrm{in}}/r_{\mathrm{out}}$ (Fig.~\ref{f02}). The set of equations solves only
for a limited range of $\alpha=0.19-0.29$, and the derived values of the others are
limited to a narrow range of $n_{\mathrm{e,in}}=10^{13.4}$--$10^{13.7}$~cm$^{-3}$ and
$f=$0.38--0.61. Within this range, $\tau_{\mathrm{es}}=$0.21--0.34 and
$N_{\mathrm{H}}=10^{23.5}$--$10^{23.7}$~cm$^{-2}$.

The WD effective area and gravity are related to the luminosity $L_{\mathrm{WD}}=0.5-0.7
\times 10^{38}$~erg~s$^{-1}$  as
\begin{eqnarray}
 L_{\mathrm{WD}} = 4\pi R_{\mathrm{WD}}^{2} \sigma_{\mathrm{SB}} T_{\mathrm{eff}}^{4},
\end{eqnarray}
and
\begin{eqnarray}
 g = G M_{\mathrm{WD}}/R_{\mathrm{WD}}^{2},
\end{eqnarray}
respectively.  Here, $R_{\mathrm{WD}}$ and $M_{\mathrm{WD}}$ are the radius and mass
of the WD, respectively, and $\sigma_{\mathrm{SB}}$ and $G$ are the Stefan-Boltzmann and
gravity constants. For $M_{\mathrm{WD}} = 1.2~M_{\odot}$, $R_{\mathrm{WD}} = 3.8 \times
10^{8}$~cm from the mass-to-radius relation \citep{nauenberg72}, which corresponds to  $\sim$0.2 
$r_{\mathrm{in}}$. Then, $T_{\mathrm{eff}} \sim 900$~kK and $\log{g} \sim 9.0$.

Some of these derived values can be verified by other relations not used above. First,
the intrinsic luminosity $L_{\mathrm{WD}}$ is consistent with the luminosity to support
steady nuclear burning on the WD surface of a mass of $\approx 1 M_{\odot}$
\citep{ebisawa01}. Second, if the observed energy shift of $\sim$900~km~s$^{-1}$
represents the radial velocity of the expanding corona:
\begin{eqnarray}
 v = \frac{\dot{M_{\mathrm{w}}}}{4 \pi r^{2} n_{\mathrm{e}} m_{\mathrm{p}}},
\end{eqnarray}
in which $\dot{M_{\mathrm{w}}}$ is the wind mass loss rate and $m_{\mathrm{p}}$ is the
proton mass. By substituting $\dot{M_{\mathrm{w}}} \sim 10^{-7} M_{\odot}$~yr$^{-1}$
\citep{ablimit2015}, we obtain $n_{\mathrm{e}}r^{2} \sim 4 \times 10^{33}$~cm$^{-1}$,
which agrees with the value derived from equation~(\ref{e1}) within an order.

\subsection{Numerical calculation}\label{s3-3}
We next verify that the analytically estimated parameters of the model (\S~\ref{s3-2})
are consistent with the numerical calculation of the photo-ionization plasma. We model
the corona using the 1D radiative transfer code \texttt{xstar} v2.58e
\citep{kallman04}. The code solves the radiative transfer in the scheme using the
two-stream (radially inward and outward directions) solver and the escape probability
approximation. It calculates the ionization/recombination, excitation/de-excitation, and
heating/cooling balance at each zone of the spherically symmetric plasma and generates
the charge state distributions and the level populations at each zone as well as
synthesized X-ray spectra inward and outward of the plasma.

The heating and cooling sources are only the radiations. The heating source, or the incident
emission, is given by the spectral energy distribution (SED) and the intensity
parameterized with $\xi$. The plasma is described by its density profile $n_{e}(r)$ and
total thickness ($N_{\mathrm{H}}$). The turbulent velocity is fixed to 400~km~s$^{-1}$
(\S~\ref{s2-3}).  We left H and He at the solar metacility and C, N, O, Ne, Mg, Si, S,
Ar, Ca, and Fe at a half solar value. We ignored other elements. The electrons have a
Maxwellian energy distribution of a temperature ($T_{\mathrm{e}}$) for the NLTE
condition, which is calculated to reach a thermal balance between the heating and
cooling.

For the purpose of a qualitative assessment, we use nearly a slab geometry of a uniform
density and a blackbody spectrum for the incident emission using the best-fit values in
the phenomenological model (\S~\ref{s2-3}). We use the code in an open geometry, in
which the emission from the photoionized plasma can escape both the inward and outward
directions, and the effects of radiative excitation are taken into account.  The
radiative excitation is known to be important for X-ray spectra from photoionized
plasmas without the incident emission in the beam such as in Seyfert 2 galaxies
\citep[e.g.,][]{kinkhabwala2002}, which should also apply to the accretion disk coronae
of X-ray binaries.

\subsubsection{Presence and absence of lines}\label{s3-3-1}
\begin{figure}[htbp]
 \begin{center}
  \includegraphics[width=1.0\columnwidth, clip]{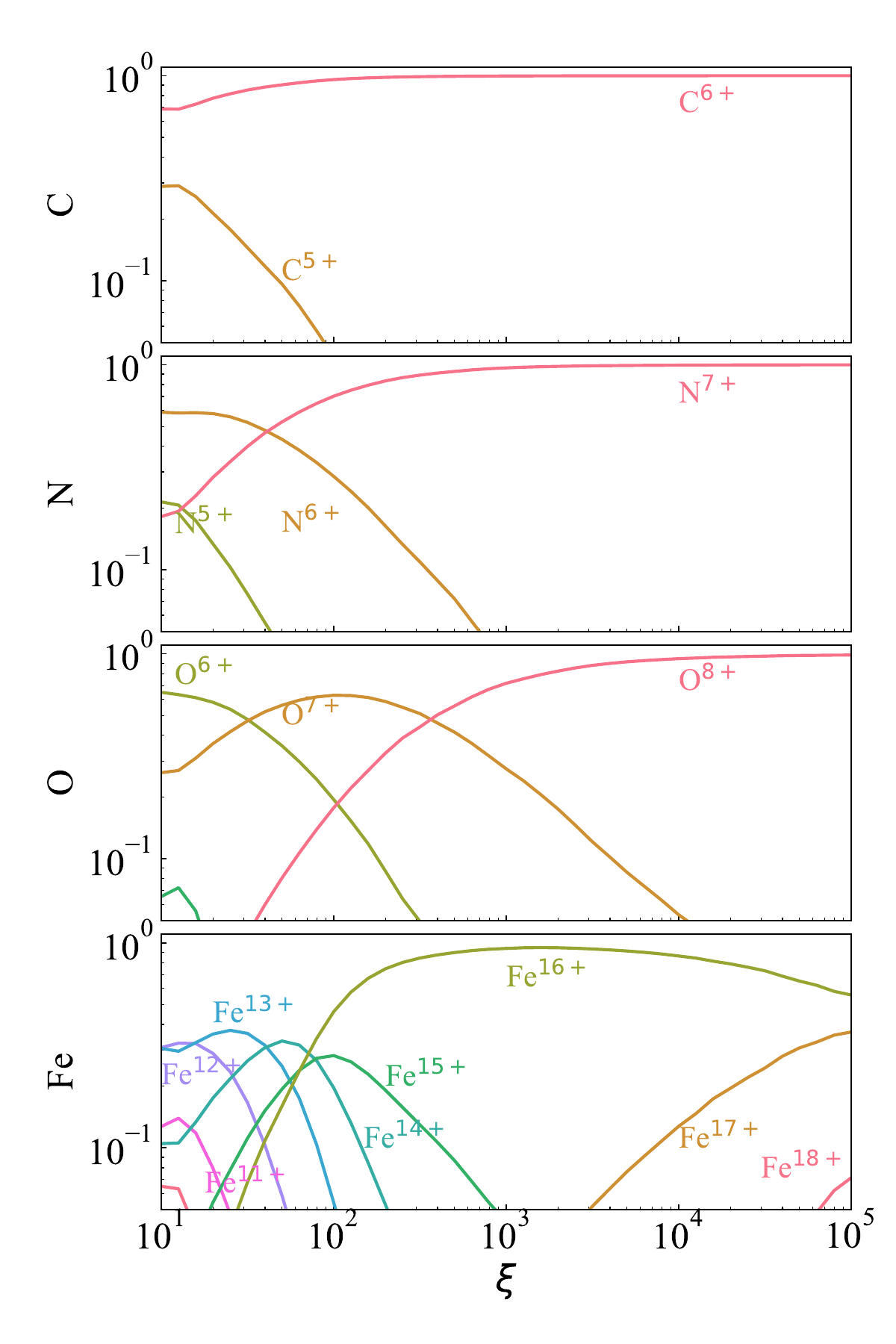}
 \end{center} 
  \caption{Charge state distribution as a function of $\log\xi$ (erg~s$^{-1}$~cm)
 calculated by \texttt{xstar} for C, N, O, and Fe averaged over the corona volume.}  
 \label{f03}
\end{figure}

We first confirm the presence or absence of several lines expected at particular
ionization states (Fig.~\ref{f05}b), which gives a constraint on the charge state
distribution and the ionization parameter $\xi$. Using parameters in the conceivable
range as $N_{\mathrm{H}}=10^{23.5}$~cm$^{-2}$ and $n_{\mathrm{e}}=10^{14}$~cm$^{-3}$, we
calculated the fraction of ionization states of C, N, O, and Fe averaged over the corona
volume (Fig.~\ref{f03}). In the observed spectrum (Fig.~\ref{f05}b), we identified
\ion{N}{7} lines (Ly$\alpha$ and Ly$\beta$) but not \ion{N}{6} lines (He$\alpha$)
produced by the recombination cascade respectively of N$^{7+}$ and N$^{6+}$ ions; these
conditions give the lower ionization limit $\log\xi \gtrsim 2.2$. For O, we identified
both \ion{O}{8} (Ly$\alpha$ and Ly$\beta$) and \ion{O}{7} features (He$\alpha$) produced
in the same process of O$^{8+}$ and O$^{7+}$ ions; these conditions in contrast gives
the upper-lmit $\log\xi \lesssim 3.5$. Over this $\xi$ range, most Fe stays in the
Ne-like (Fe$^{16+}$), which is consistent with the presence of \ion{Fe}{17} features
produced by radiative decay from excited levels to the ground state of Fe$^{16+}$
ions. The absence of \ion{Fe}{16} and \ion{Fe}{18} features is due to the small fraction
of Fe$^{15+}$ and Fe$^{17+}$ ions, respectively. These conditions are consistent with
the estimated $\xi$ range of $10^{2.2}-10^{3.5}$.

\subsubsection{Line ratios}\label{s3-3-2}
We next examine the intensity ratio of the identified lines.  A well-established method
is to use the He$\alpha$ triplet line intensity ratios of \ion{O}{7}, namely the
resonance ($r$), inter-combination ($i$), and forbidden lines ($f$).  $G \equiv (f+i)/r$
and $R \equiv f/i$ ratios are often used as temperature and density indicators when
collisional processes dominate \citep{gabriel69} or the ionization and column density
indicators when radiative processes dominate \citep{porter2007}. We refrain from using
these diagnostics to constrain model parameters, though, given the poor constraints by
the data (Fig.~\ref{f05}a inset). We only argue that it is reasonable to have the $r$
line to be the strongest and the $f$ line to be the weakest. The reasons for the former
are that (1) $T_{\mathrm{e}}=0.2-2.0$~MK from the thermal balance is close to the
maximum formation temperature of the line in the collisional process and (2) the line is
enhanced by the continuum pumping in the radiative process \citep{Chakraborty2021}. The
reason for the latter is that $n_{\mathrm{e}} \sim 10^{14}$~cm$^{-3}$ (\S~\ref{s3-2}) is
much higher than the critical density of $3.4 \times 10^{10}$~cm$^{-3}$ for O
\citep{Blumenthal1972}.

\section{Physical Modeling}\label{s4}
\begin{figure*}[p]
 \begin{center}
  \includegraphics[width=0.9\textwidth]{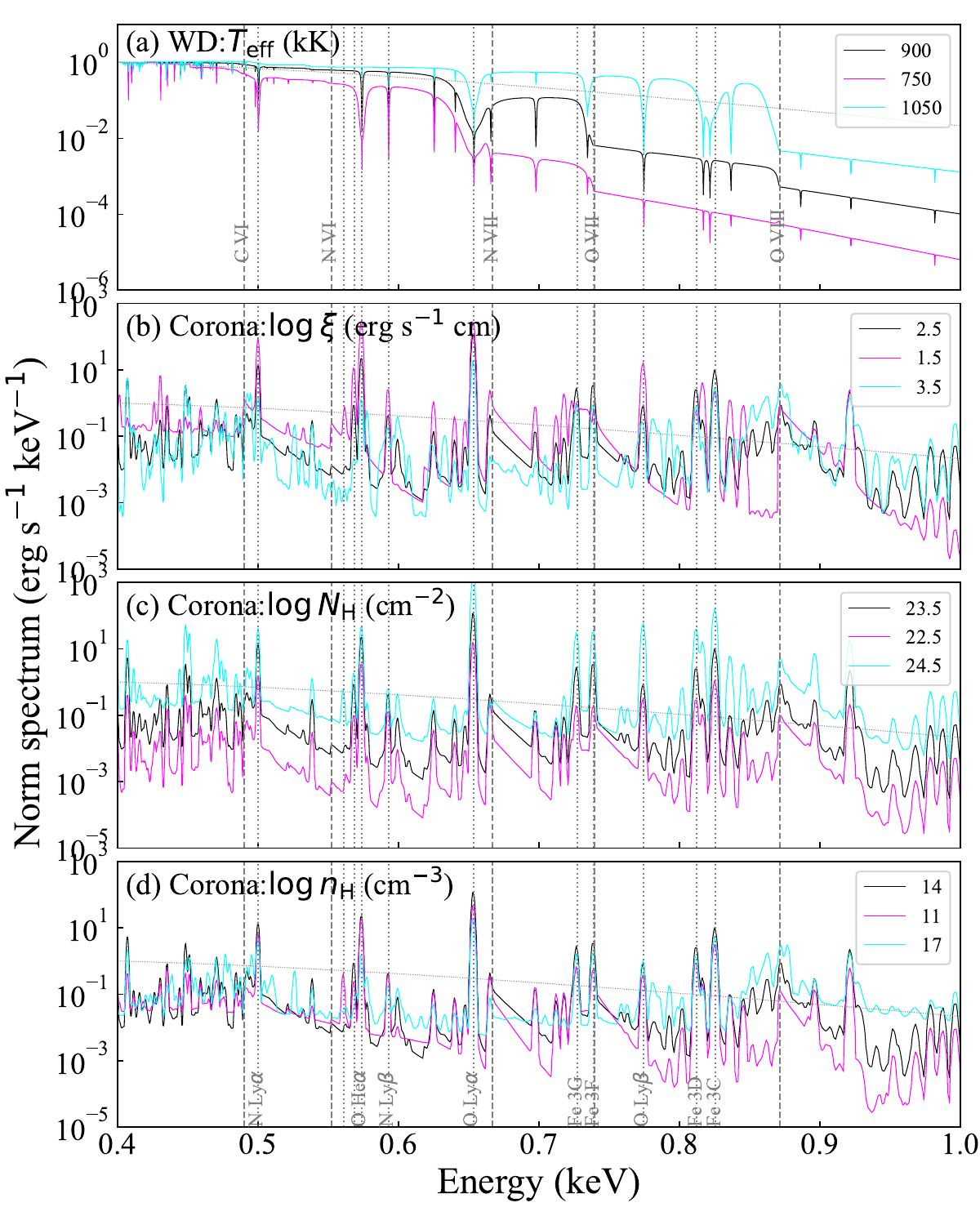}
  \caption{Synthesized spectra of (a) incident emission from the WD atmosphere and
  (b--d) diffuse emission from the corona with different physical parameters.  Note that
  the incident emission for the corona model is a blackbody emission shown with the
  dotted grey curve in (a--d). Spectra in (a) are normalized at 0.4~keV. For each of the
  model parameters ($T_{\mathrm{eff}}$, $\log \xi$, $\log N_{\mathrm{H}}$, and $\log
  n_{\mathrm{H}}$), three representative values are chosen; one is a value close to the
  estimates (\S~\ref{s3}) in black (thus the black line spectra in the panels (b-d) are
  identical) and a lower (magenta) or a higher (cyan) value than that. Absorption edges
  are labeled in (a) and shown with dashed vertical lines. Lines are labeled in (d) and
  shown with dotted vertical lines.}
 \label{f04}
 \end{center}
\end{figure*}

Based on the comparison of the observed spectra (\S~\ref{s2-3}) against the analytical
(\S~\ref{s3-2}) and numerical (\S~\ref{s3-3}) calculations, we have verified that our
assumed model and the parameterization are appropriate.  We have also constrained the
range of parameters. In order to contain the parameters more accurately, we now
construct physical models using two complementary methods: one is 1D two-steam solver
(\texttt{xstar}; \S~\ref{s4-1}) and the other is 3D Monte Carlo solver (\texttt{MONACO};
\S~\ref{s4-2}) of the radiative transfer in the corona.

\subsection{1D two-stream solver}\label{s4-1}
\subsubsection{Setup}\label{s4-1-1}
For the corona model, we use \texttt{xstar} in the same setup with \S~\ref{s3-3} except
for two improvements; inclusion of the density profile $n_{\mathrm{e}}(r) =
n_{\mathrm{e,in}} \left(r/r_{\mathrm{in}}\right)^{-2}$ and  use of a WD atmosphere
model spectrum for the incident emission. For the atmosphere model, we used
\texttt{TMAP} \citep{rauch2010}, which provides spectra\footnote{The data are available
at \url{http://astro.uni-tuebingen.de/~rauch/TMAF/flux_HHeCNONeMgSiS_gen.html}.} at
10---8000~\AA\ for $T_{\mathrm{eff}}=$405--1050~kK and $\log{g}$~(cm~s$^{-2}$)$=$9 with
varying abundances for major elements.

Figure~\ref{f04} shows the synthesized spectra for different (a) atmosphere and (b--d)
corona model parameters. Note that the incident emission for the corona model (b--d) is
a featureless blackbody emission to clarify the corona effects, so they should be
considered a transfer function. For $T_{\mathrm{eff}}$, the most conspicuous changes are
found in the \ion{O}{7} and \ion{O}{8} edges. Their depth ratio is sensitive to
$T_{\mathrm{eff}}$. The edges, however, are insignificant in the corona transfer
function in (b--d) as they are overwhelmed by the radiative recombination continuum
emission.  For $\log{\xi}$, the charge state distribution changes significantly
(Fig.~\ref{f03}) and, resultantly, the emission line ratio of different charge states;
e.g., \ion{O}{8} Ly$\alpha$ and \ion{O}{7} He$\alpha$. For $\log{N_{\mathrm{H}}}$, the
emission line ratios are not influenced, but the spectral shape above the edge energies
change due to different columns in the line of sight. For $\log{n_{\mathrm{H}}}$, the
\ion{O}{7} triplet changes most significantly (\S~\ref{s3-3-2}).

\subsubsection{Result}\label{s4-1-2}
We ran the \texttt{xstar} calculations for a grid of three parameters:
$T_{\mathrm{eff}}$, $\xi$, and $N_{\mathrm{H}}$. We skip $n_{\mathrm{H}}$ as the
constraining power of the \ion{O}{7} triplet in the data is poor. The Galactic and
Magellanic interstellar absorption columns were fixed to the value derived in the
phenomenological fitting in Table~\ref{t01}. The turbulent velocity is fixed to
400~km~s$^{-1}$ (\S~\ref{s2-3}). The redshift ($v_{\mathrm{shift}}$) was fitted
collectively. All the other parameters, including $n_{\mathrm{H}}$, are derived from the
free parameter values based on the equations in \S~\ref{s3-2}.

Figure~\ref{f12} (a) shows the best-fit model and the residuals to the fit. The fitting
is very poor. The best-fit parameters are away from those in the analytical estimates
(\S~\ref{s3}). We identify several major discrepancies: emission lines are too strong,
continuum emission is too weak, and the line profiles and ratios are poorly
reproduced. Many of these are expected due to the known limitations of this solver,
which we will discuss in \S~\ref{s5}.

\subsection{3D Monte Carlo solver}\label{s4-2}
\subsubsection{Setup}\label{s4-2-1}
We next use the other radiative transfer code \texttt{MONACO} based on the Monte Carlo
solver. \texttt{MONACO} is a photon tracking simulator based on the Monte Carlo
framework \citep{watanabe2006,odaka2011,odakaDthesis} and is applied to many
astrophysical applications (e.g.,
\citealt{hagino2015,tomaru2018,tanimoto2019,mizumoto21}). We used \texttt{MONACO}
version 1.7 release candidate and the database version 1.7.

A number of photons are emitted from the source and individual photons are tracked for
their interactions with the matter placed in a 3D space. Each photon is tracked until it
either escapes from the system or is destructed inside the system by
absorption. Interactions between the photons and matter are calculated, including
radiative excitation/de-excitation, ionization/recombination, photoelectric absorption,
and scattering by free and bound electrons. However, because the framework is based on
photon tracking, emission processes initiated by electrons such as free-free emission,
are currently not included.  We collected photons that experienced at least one
interaction in the corona for the secondary emission, which includes both the scattered
and diffuse emission.

The design of the framework allows one to calculate electron scattering in a 3D space in
a consistent manner. The code does not calculate the thermal, ionization, and excitation
balances unlike \texttt{xstar}. The electron density and temperature ($n_{\mathrm{e}}$
and $T_{\mathrm{e}}$) as well as the charge state distributions of ions at each location
need to be given as inputs. We used the result of the \texttt{xstar} simulation with the
best-fit parameters (\S~\ref{s4-1}).

In the current version, \texttt{MONACO} calculates the level populations determined
purely by the collisional processes based on the given $n_{\mathrm{e}}$ and
$T_{\mathrm{e}}$. This is not correct for the photo-ionized plasmas, in which the
radiative processes also contribute. Unlike \texttt{SKIRT} version 9.0, a widely used
Monte Carlo solver \citep{vandermeulen2023}, \texttt{MONACO} calculates the photon-ion
interactions for the H-like and He-like ions of major metals. For this work, we made two
modifications to the code and the database: (1) inclusion of lower ionized Fe ions than
Fe$^{24+}$ down to Fe$^{16+}$ and (2) recording both the positive and the negative
deposit energy for the resonance scattering.

\subsubsection{Result}\label{s4-2-2}
\begin{figure*}
 \begin{center}
  \includegraphics[width=1.0\columnwidth]{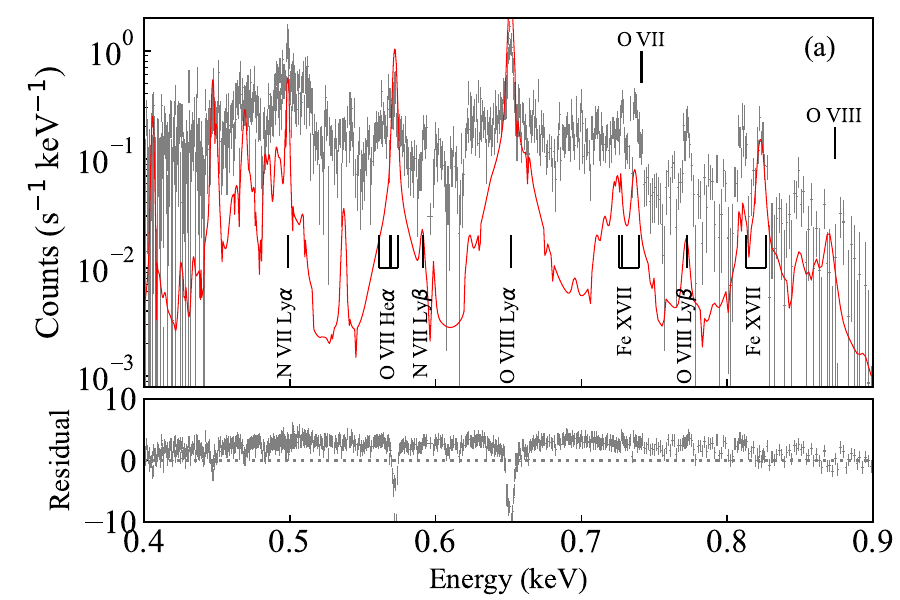}
  \includegraphics[width=1.0\columnwidth]{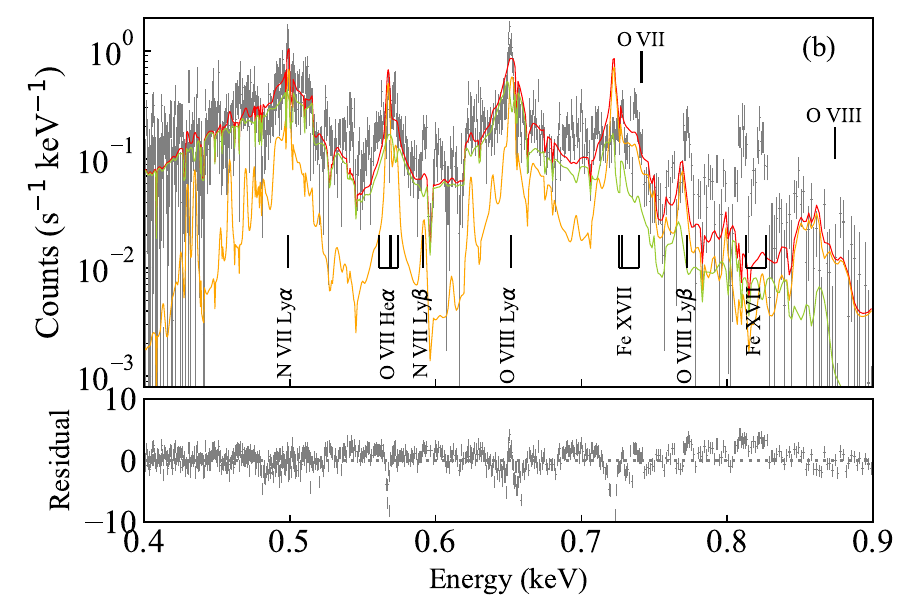}
  \caption{Fitting result with the physical model based on (a) \texttt{xstar} and (b)
  \texttt{MONACO}. Symbols follow Figure~\ref{f05}. In (b), the model spectrum is
  decomposed into the scattered emission (yellow-green) and the diffuse emission
  (orange) based on the last interaction of the photons escaping from the system.}
 \label{f12}
 \end{center}
\end{figure*}

We used the same approach with the 1D two-stream solver (\S~\ref{s4-1}) for the
fitting. The geometry (Fig.~\ref{f01}) is set up in a 3D space. The simulation was run
at the same grid of the same free parameters ($T_{\mathrm{eff}}$, $\xi$, and
$N_{\mathrm{H}}$) with the turbulent velocity fixed, the bulk velocity collectively
fitted, other corona parameters derived from the grid parameters based on the analytical
relations (\S~\ref{s3-2}), and the Magellanic and Galactic absorption fixed to the value
from the phenomenological fitting.

For each grid parameter, we launched half a million photons from a point source at the
center. The photons have a direction selected randomly over a uniform distribution and
an energy selected randomly over a uniform distribution in 0.35--1.24~keV. The output
was convolved with the input SED. For the grid of $\log{\xi}=2.5$,
$\log{N_{\mathrm{H}}}=23.5$, and $T_{\mathrm{eff}}=10^{3}$~kK as an example, (a) 4\%
photons are destructed in the corona by photo-electric absorption, (b) 72\% photons
escape from the system with no interactions, and (c) the rest escape with some
interactions in the corona. We use (c) for the scattered plus diffuse emission.

Figure~\ref{f12} (b) shows the best-fit model and the residuals to the fit. The fitting
has improved with the best-fit parameters of $\log{\xi}=2.6$,
$\log{N_{\mathrm{H}}}=23.2$, and $T_{\mathrm{eff}}=1015$~kK. They are in reasonable
agreement with the analytical solution, yet the fitting is still too poor to derive
their uncertainties. Some discrepancies remain, in particular, in the line profile of
\ion{O}{8} Ly$\alpha$, the line ratio within \ion{O}{7} He$\alpha$, and \ion{Fe}{17}
lines. We will discuss these issues in \S~\ref{s5}.

\section{Discussion}\label{s5}
We constructed the spectral models of the corona based on two different solvers of the
radiative transfer calculation and compared them to the observed spectra (\S~\ref{s4}). The
\texttt{xstar} model yielded a poor fitting with the best-fit parameters far from those
by the analytical estimates (\S~\ref{s3-2}). The \texttt{MONACO} model yielded a better
fitting with the best-fit parameters consistent with the analytical estimates
(\S~\ref{s3-2}). Both of them exhibited some discrepancies against the observations. We
discuss possible reasons for four major discrepancies (\S~\ref{s5-1}--\S~\ref{s5-4}) and
present the final fitting by considering the model limitations (\S~\ref{s5-5}).

\subsection{Continuum emission}\label{s5-1}
A major discrepancy in the \texttt{xstar} fitting (Fig.~\ref{f12} a) is too weak
continuum emission in the model. This makes sense considering that most of the
continuum emission comes from the scattered emission by electrons in the corona, not the
diffuse emission. In the 1D calculation of \texttt{xstar}, the electron scattering is
implemented as a pure continuum absorption, and no scattered emission is produced.

In \texttt{MONACO}, the electron scattering is considered as scattering, including
multiple scattering in the 3D space. We decompose the \texttt{MONACO} model spectrum
into the scattered and diffuse components (Fig.~\ref{f12} b). A single photon cannot be
attributable to either one of them; a photon can experience both electron scattering and
resonance scattering before escaping from the system. Here, the resonance scattering is
not scattering but is absorption immediately followed by emission, thus should be
included in the diffuse component. We divided all the photons based on their last
interactions in the system. As expected, the scattered emission accounts for a large
part of the continuum emission.

\subsection{Line intensity}\label{s5-2}
\begin{figure*}
 \centering
 \includegraphics[width=1.0\textwidth,clip]{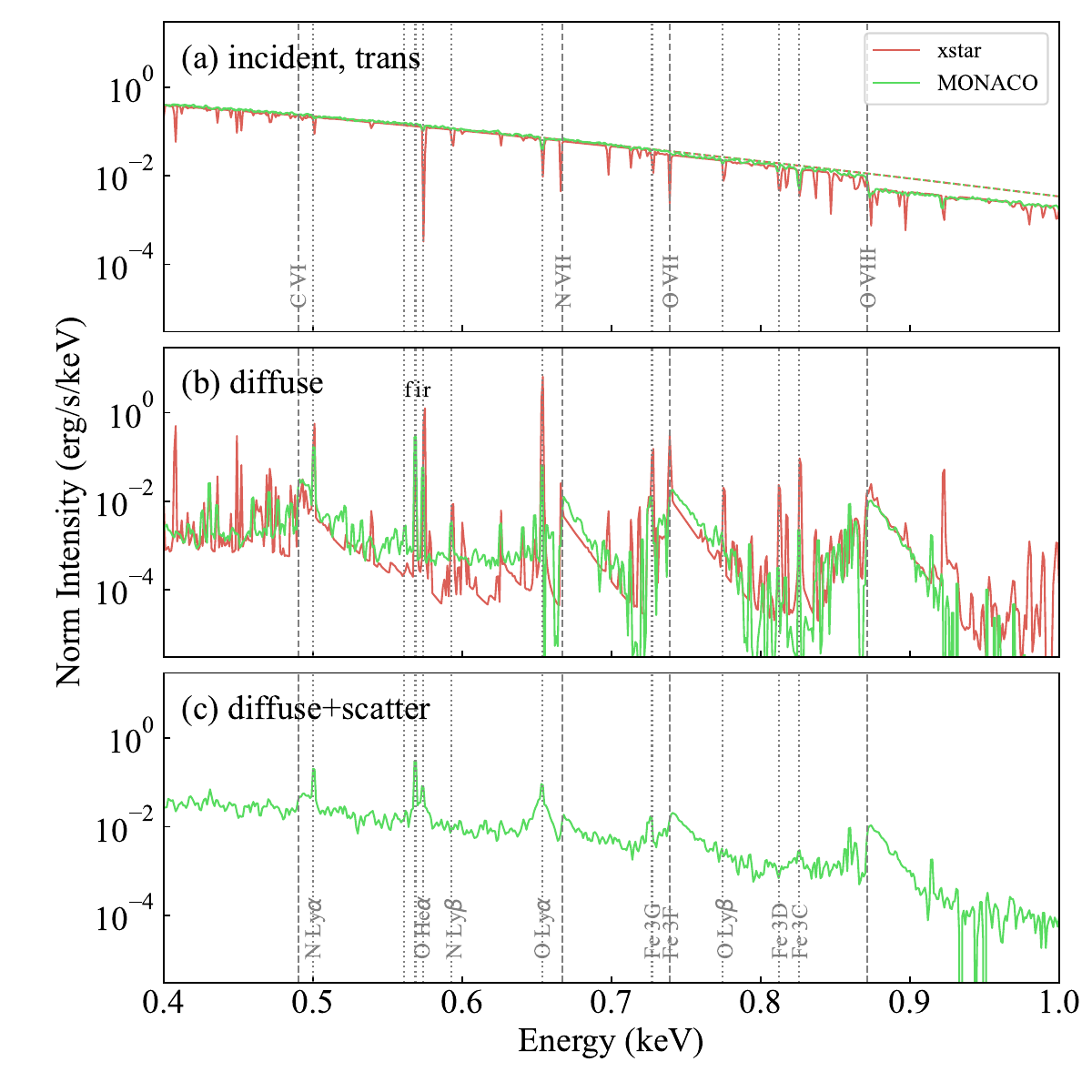}
 \caption{Comparison between the \texttt{xstar} and \texttt{MONACO} model spectra for
 the same setup (80~eV blackbody emission for the input with
 $\xi=10^{2.5}$~erg~s$^{-1}$~cm and $N_{\mathrm{H}}=10^{22.5}$~cm$^{-2}$). (a) Incident
 (dashed) and transmitted (solid) emission, (b) diffuse emission, and (c) diffuse and
 scattered emission combined, which is available only for \texttt{MONACO}.
 }
 \label{f20a}
\end{figure*}

Another discrepancy in the \texttt{xstar} fitting (Fig.~\ref{f12}a) is the too strong
line emission. The lack of scattered emission in the model accounts for this
partially, but not fully. To illustrate this, we compare the \texttt{xstar} and
\texttt{MONACO} spectra for the same parameters in Figure~\ref{f20a}. For the
transmitted spectra in panel (a), the continuum levels agree well between the two.  For
the diffuse spectra in panel (b), the radiative recombination continua at \ion{C}{6},
\ion{N}{7}, \ion{O}{7}, and \ion{O}{8} agree in general. However, the line emission is
systematically stronger for \texttt{xstar} than \texttt{MONACO}.

We argue two possibilities for the difference between the two solvers. One is that
\texttt{MONACO} assumes that all excitation occurs only collisionally and does not
include radiative excitation, which is known to enhance the emission lines
\citep{Chakraborty2021}. The other is the assumption on the escape probability employed
in \texttt{xstar}. The escape probability is an assumption necessary to solve the
radiative transfer equations in reasonable computational resources in two-stream
solvers, in which line photons escape from the system at a probability of
$e^{-\tau(\nu)}$. Here, $\tau(\nu)$ is the opacity in the direction of the photon with a
frequency of $\nu$. It has been claimed that solvers based on the escape probability
assumption yield line emission stronger than the accelerated lambda iteration solver,
which is considered more reliable for the lines \citep{hubeny01}. In fact,
\citet{dumont2003} argued that the escape probability method yields an overestimation of
the line intensity, in particular for resonance lines like \ion{O}{8} Ly$\alpha$ in
optically thick cases, by an order.

\subsection{Line profile}\label{s5-3}
\begin{figure}
 \begin{center}
  \includegraphics[width=1.0\columnwidth]{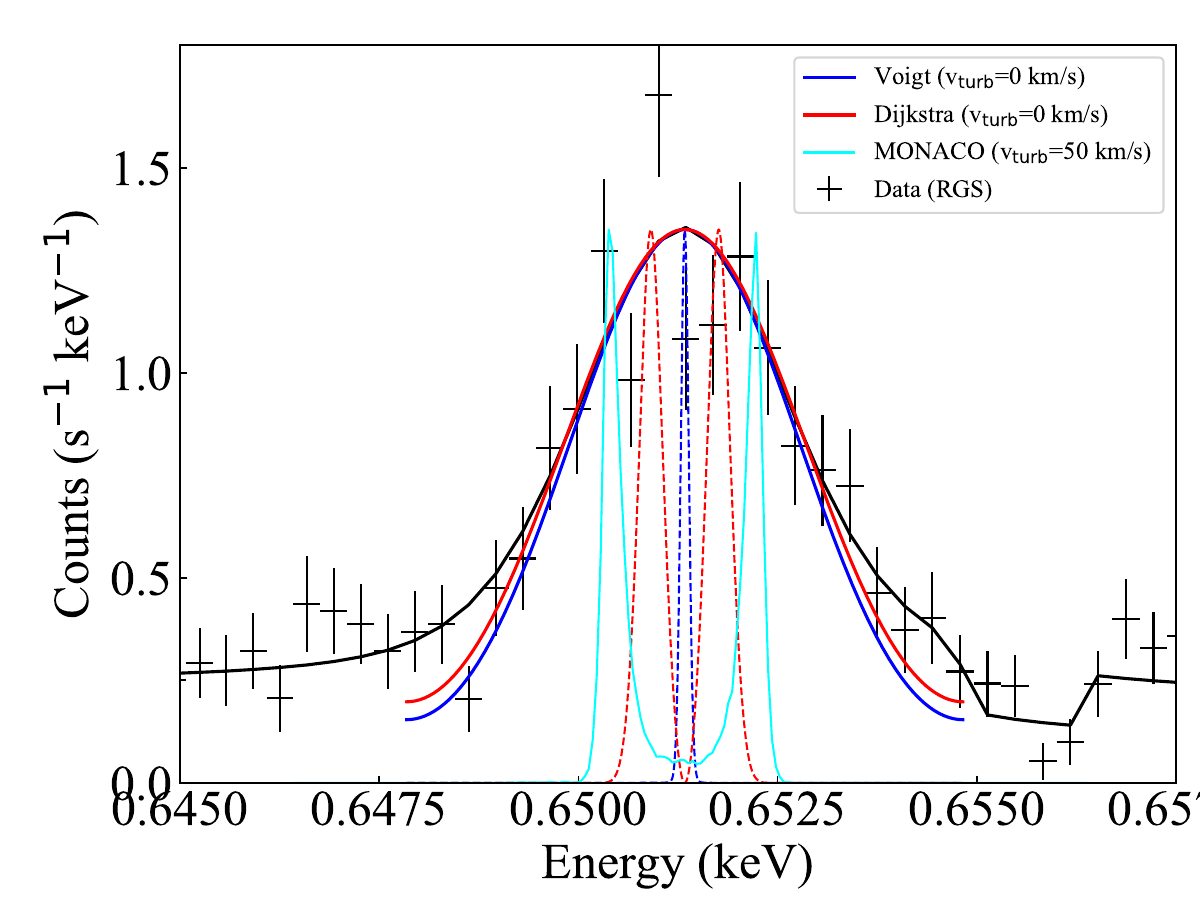}
  \caption{Profile of the \ion{O}{8} Ly$\alpha$ line. The data are shown with black
  error bars, while the best-fit phenomenological model (\S~\ref{s2-3}) is shown with
  the black curve. The phenomenological model has two unresolved Gaussian lines
  representing Ly$\alpha ^{1}$ and Ly$\alpha ^{2}$ with a fixed energy and intensity
  ratios, which are collectively shifted redward by 1$\times$10$^{3}$~km~s$^{-1}$ and
  smoothed by 360~km~s$^{-1}$ (Fig.~\ref{f10}) and convolved with the detector line
  spread function (LSF). Upon this, compared are the Voigt (blue) and Harrington (red)
  profiles. The profiles are broadened only for the natural and the thermal broadening
  of 10$^{6}$ K and not by turbulence. The compared profiles are convolved with the
  Gaussian core of the LSF but no underlying continuum emission, thus the tail parts
  should be ignored. The cyan curve shows the profile calculated by \texttt{MONACO} with
  finite but negligible turbulence (50~km~s$^{-1}$) to stabilize the calculation.
  }
 \label{f18}
 \end{center}
\end{figure}

The opacity at the line center is very large for the conspicuous lines. Taking the
strongest line \ion{O}{8} Ly$\alpha$ as an example, it is estimated by
\begin{equation}
 \label{e09}
 \tau_{\mathrm{O~Ly}\alpha} = \frac{1}{4\pi\epsilon_0} \frac{\pi e^{2}}{m_{\mathrm{e}}c} gf
 \frac{1}{\sqrt{\pi}\Delta\nu_{\mathrm{D}}}
 N_{\mathrm{H}} A_{\mathrm{O}} A_{8+} A_{\mathrm{1s}},
\end{equation}
in which
$\Delta\nu_{\mathrm{D}}=\frac{E_0}{hc}\sqrt{\frac{2k_{\mathrm{B}}T}{m_{\mathrm{O}}}}$,
$E_0$ is the energy of the line, $gf$ is the weighted oscillator strength,
$A_{\mathrm{O}}$ is the oxygen abundance relative to hydrogen, $A_{8+}$ is the charge
fraction of O$^{8+}$, $A_{\mathrm{1s}}$ is the level population fraction of the ground
state, $m_{\mathrm{O}}$ is the Oxygen atomic mass, $T$ is the plasma temperature, and
other symbols follow their conventions. For $E_0=653$~eV, $f=0.14$,
$N_{\mathrm{H}}=10^{23.5}$~cm$^{-2}$, $A_{\mathrm{O}}=0.5 \times 4.9 \times 10^{-4}$,
$A_{8+}=0.5$ (Fig.~\ref{f03}), $A_{\mathrm{1s}} = 1$, and $T=10^{6}$~K, we obtain
$\tau_{\mathrm{Ly}\alpha} \sim 6.7 \times 10^{4} \; \tau_{\mathrm{es}} \sim 1.4 \times
10^{4}$.

Line photons scatter numerous times locally due to their large opacity at the line
center until their wavelength diffuses into the dumping wing of the line profile. Then,
the photons escape from the system in a single long flight with a significantly reduced
opacity. This is a premise for the escape probability approximation implemented in
\texttt{xstar}. In such a situation, the line profile is known to be double-peaked
around the center. The analytical solution is obtained for a plane-parallel
\citep{harrington1973,neufeld1991} and spherical \citep{dijkstra2006} geometry of a
uniform density. Despite the premise, \texttt{xstar} synthesizes emission lines with the
Voigt profile as it is computationally challenging to calculate the diffusion over
wavelengths.

Figure~\ref{f18} shows a close-up view of the \ion{O}{8} Ly$\alpha$ profile. The Voigt
and Dijkstra profiles of no turbulence are shown with red and blue broken curves. They
are smoothed with a Gaussian of the RGS energy resolution to compare to the data as
shown with the solid curves of the same color. The smoothed Dijkstra profile matches
better to the data than the smoothed Voigt profile. The turbulence velocity required in
the phenomenological fitting (Fig.~\ref{f10}) can be explained by the line profile
distortion by the radiative transfer effects. \texttt{MONACO} calculates the energy
shift in every resonance scattering, thus the emergent line profile exhibits a
distortion as shown with a cyan curve in Figure~\ref{f10}. This is too broad compared to
the data, which we come back to later in \S~\ref{s5-5}.

\subsection{Line ratios}\label{s5-4} 
We discuss discrepancies found in several line ratios. First is the \ion{O}{7}
He$\alpha$ triplet. The line ratios are different between \texttt{xstar} and
\texttt{MONACO} most notably in their synthesized diffuse spectra (Fig.~\ref{f20a}
b). Among the resonance ($r$), inter-combination ($i$), and forbidden ($f$) lines, $f$
is the weakest in both models, which is expected (\S~\ref{s3-3-2}) and agrees with the
observation (Fig.~\ref{f05}). However, the $i$/$r$ ratio in the \texttt{MONACO} model is
too strong compared to \texttt{xstar} and the observation. This stems from the
limitation of \texttt{MONACO}, in which the level populations are calculated assuming
that they are purely dominated by collisional processes. In the present application,
both collisional and radiative processes matter, which are taken into account in the
\texttt{xstar} calculation.

For the line ratios, we should focus on the \texttt{xstar} results. We examine the
\ion{O}{8} Lyman decrement (Ly$\beta$/Ly$\alpha$). The observed decrement is $\sim$1/3
(Fig.~\ref{f10}), while the calculated value is $\sim$0.05 (Fig.~\ref{f04}). The
observed Ly$\beta$ is too strong with respect to Ly$\alpha$. The calculated decrement
hardly changes if we change the model parameters. Another line ratio discrepancy is
found in the four \ion{Fe}{17} lines (Fig.~\ref{f12} a), in which the observed 3C/3D
line pair is too strong with respect to the 3F/3G pair. Both are primarily due to the
attenuation of incident photons beyond the \ion{O}{7} edge.

A possible solution would thus be to introduce another spectral component such as the
collisionally ionized emission independent of the \ion{O}{7} edge attenuation. A
plausible origin of the collisionally ionized plasma is the shock of the expanding shell. If
we assume that it is produced by the shock of the expanding velocity of $v \sim
900$~km~s$^{-1}$ (Fig.~\ref{f10}), the temperature is given by
\begin{math}
 T = \frac{3}{16}\frac{\mu m_{\mathrm{p}}}{k_{\mathrm{B}}} v^{2} = 1.0~\mathrm{keV},
\end{math}
in which $\mu$ is the mean molecular weight, $m_{\mathrm{p}}$ is the proton mass, and
$k_{\mathrm{B}}$ is the Boltzmann constant. Such a spectral component has been found in
other SSS \citep{ness2022}.

\subsection{Final fitting}\label{s5-5}
\begin{figure}
 \begin{center}
  \includegraphics[width=1.0\columnwidth]{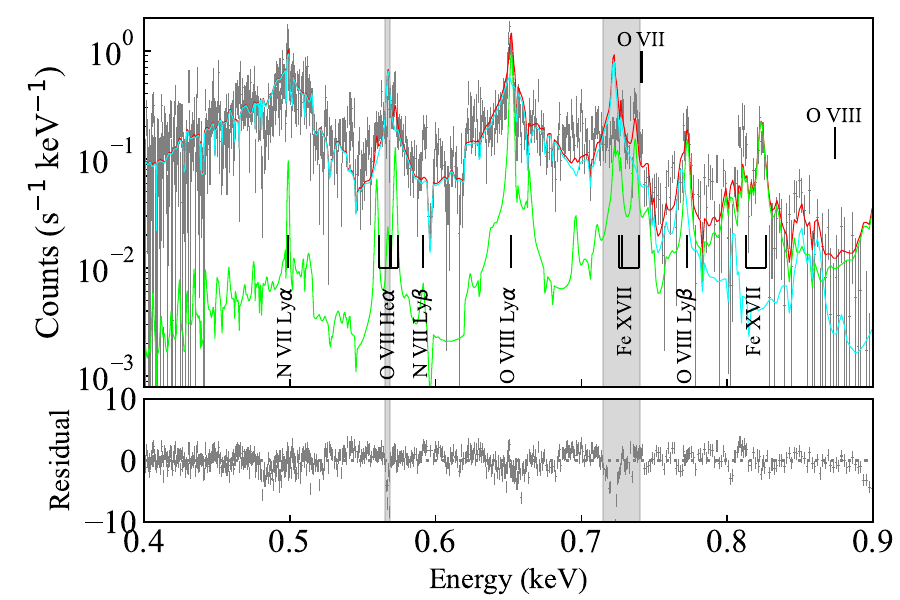}
  \caption{Final fitting using the corona model of \texttt{MONACO} (cyan) plus a
  collisionally ionized plasma model of \texttt{apec} (lime). The energy range of
  0.565--0.569 and 0.715-0.740~keV are ignored to avoid the known caveats of the
  \texttt{MONACO} modeling for the \ion{O}{7} He$\alpha$ ($i$) and \ion{Fe}{17} 3F/3G
  lines.  }
 \label{f21}
 \end{center}
\end{figure}

Based on the discussion above, we modify the spectral model for the final fitting. We
start with the \texttt{MONACO} model (Fig.~\ref{f12}b) and add an \texttt{apec} model to
account for the additional collisionally ionized plasma. We ignore the energy range of
0.565--0.569 and 0.715-0.740~keV to avoid the known caveats of the \texttt{MONACO}
modeling for the \ion{O}{7} He$\alpha$ ($i$) and \ion{Fe}{17} 3F/3G lines.  The plasma
temperature ($T$) and the normalization of the \texttt{apec} model were fitted, while
the abundance was fixed to half the solar value. Both the \texttt{MONACO} and
\texttt{apec} models were red-shifted collectively. The result is given in
Figure~\ref{f21}.

The fitting is finally good enough (reduced $\chi^2=$1.5) to derive the best-fit
parameter ranges as $\log{\xi}$ (erg~s$^{-1}$~cm) $=$2.52--2.55, $\log{N_{\mathrm{H}}}$
(cm$^{-2}$) $=$23.49--23.50, $T_{\mathrm{eff}}$ (kK) $=$1000--1005,
$v_{\mathrm{redshift}}$ (km~s$^{-1}$) $=$ 895--941, and $T$ (keV) $=$ 0.24--0.25. The
corona model parameters agree well with the analytical solution. The \ion{O}{8} Lyman
decrement is now better reproduced since the Ly$\beta$ is contributed by the
\texttt{apec} component. The \ion{Fe}{17} 3C line is also reproduced by the
\texttt{apec} model, which was not accounted for in the corona model only with
\texttt{MONACO}. The \ion{O}{8} Ly$\alpha$ profile was better reproduced by a
combination of the broad corona profile and the narrow collisionally ionized plasma
profile. We still see some discrepancies such as \ion{Fe}{17} 3D line at 0.81~keV and
\ion{C}{6} radiative recombination continuum at 0.49~keV. These, along with the
identified caveats and justification of the additional \texttt{apec} model, are
left for future improvements of the codes.

\section{Summary and Conclusion}\label{s6}
We presented result of the spectral modeling of CAL87, a super-soft source dominated by
emission features originating from the extended accretion disk corona around the
WD. First, we presented a phenomenological fitting of the observed spectra with the
EPIC-MOS and RGS instruments onboard \textit{XMM-Newton}. Based on the fitting results
and the system geometry, an analytical solution was obtained.

We then performed radiative transfer calculations using two codes employing different
solvers and assumptions: \texttt{xstar} for a 1D, two-streamer solver and
\texttt{MONACO} for a 3D, Monte Carlo solver. We fitted the spectral model to the data
and identified discrepancies between the codes against the observation.

We discussed possible causes of these discrepancies in terms of the continuum emission,
line intensity, line profile, and line ratio. We found that \texttt{xstar} excels in the
level population calculations, while \texttt{MONACO} excels in the scattering (both
electron and resonant line) calculations. We also argued for the presence of the
additional collisionally ionized plasma. Finally, we presented the detailed spectral
model that yields consistent parameters with the analytical model and agrees reasonably
well with the observation.

\medskip

In the coming microcalorimetry era in X-ray astronomy, the interpretation based on
radiative transfer will be more seriously required than before. This is
particularly the case for sources with hard X-ray spectrum dominated by reprocessed
emission in the surrounding medium such as X-ray binaries during eclipse, accretion disk
corona sources, and Seyfert 2 galaxies. Comparisons between different solvers and
implementations, and comparisons against observations should be made in a wider
range of applications than the one presented here.

\begin{acknowledgments}
 We are grateful for our reviewer Jan-Uwe Ness for his insightful comments that
 not just improved the manuscript but gave us inspirations for the present work.
 We appreciate expert advice from Timothy Kallman at NASA GSFC for the implementation
 in \texttt{xstar} and Atsushi Tanimoto at Kagoshima University in \texttt{MONACO}. This
 work made use of the JAXA's super-computing system JSS3. This work was supported by
 JSPS KAKENHI Grant Numbers JP14J11810, JP19H01906, JP19H05185, JP18H05861, JP21K13958,
 and JP24740190. 
\end{acknowledgments}
\vspace{5mm}
\facility{
\textit{XMM-Newton} \citep{jansen2001} 
(
EPIC-MOS \citep{Turner2001}, 
RGS \citep{denherder2001}
)
}
\software{
\texttt{SAS},
\texttt{Xspec} \citep{arnaud1996},
\texttt{xstar} \citep{kallman04},
\texttt{MONACO} \citep{watanabe06,odaka11,odakaDthesis},
}
\bibliography{main}{}
\bibliographystyle{aasjournal}

\end{document}